\documentclass[twocolumn]{aa}      

\usepackage{natbib}
\usepackage{graphicx}
\usepackage{txfonts}
\usepackage[draft]{hyperref}

\begin{document} 
  \title{Discovery of a point-like source and a third spiral arm in the transition disk around the Herbig Ae star MWC~758}
\titlerunning{A point-like source and a third spiral arm in the transition disk around MWC~758}

\author{M. Reggiani\inst{1}
     \and V. Christiaens\inst{1,2,3}
     \and O. Absil\inst{1}\thanks{F.R.S.-FNRS Research Associate}
     \and D.~Mawet\inst{4,5}
     \and E. Huby\inst{1,6}
     \and E.~Choquet\inst{5}\thanks{Hubble Fellow}
     \and C.~A.~Gomez Gonzalez\inst{1}
     \and G.~Ruane\inst{4}
     \and B.~Femenia\inst{7}
     \and E.~Serabyn\inst{5}
     \and K.~Matthews\inst{4}
     \and M.~Barraza\inst{2,3}
     \and B.~Carlomagno\inst{1}
     \and D.~Defr\`ere\inst{1}
     \and C.~Delacroix\inst{1,8} 
     \and S.~Habraken\inst{1}
     \and A.~Jolivet\inst{1}
     \and M.~Karlsson\inst{9}
     \and G.~Orban de Xivry\inst{1}
     \and P.~Piron\inst{9}
     \and J.~Surdej\inst{1}
     \and E.~Vargas Catalan\inst{9}
     \and O.~Wertz\inst{10}
          }

  \institute{Space sciences, Technologies and Astrophysics Research (STAR) Institute, Universit\'{e} de Li\`ege, 19 All\'{e}e du Six Ao\^ut, B-4000 Li\`ege, Belgium\\            \email{mreggiani@ulg.ac.be}
  \and Departamento de Astronom\'{i}a, Universidad de Chile, Casilla 36-D, Santiago, Chile
  \and Millenium Nucleus "Protoplanetary Disks in ALMA Early Science", Chile
  \and Department of Astronomy, California Institute of Technology, 1200 East California Boulevard, MC 249-17, Pasadena, CA 91125 USA
  \and Jet Propulsion Laboratory, California Institute of Technology, 4800 Oak Grove Drive, Pasadena, CA 91109, USA
  \and LESIA, Observatoire de Paris, PSL Research University, CNRS, Sorbonne Universit\'{e}s, UPMC Univ. Paris 06, Univ. Paris Diderot, Sorbonne Paris Cit\'{e}, 5 place Jules Janssen, 92195 Meudon, France
  \and W. M. Keck Observatory, 65-1120 Mamalahoa Hwy., Kamuela, HI 96743, USA
  \and Mechanical \& Aerospace Engineering, Princeton University, Princeton, NJ 08544, USA
  \and Department of Engineering Sciences, \r{A}ngstr\"{o}m Laboratory, Uppsala University, Box 534, SE-751 21 Uppsala, Sweden
  \and Argelander-Institut f\"{u}r Astronomie, Auf dem H\"{u}gel 71, D-53121 Bonn, Germany}

  \date{Received 29 September 2017; after first revision with A\&A}

  \abstract
   {Transition disks offer the extraordinary opportunity to look for newly born planets  and investigate the early stages of planet formation.}
   {In this context we observed the Herbig A5 star MWC~758 with the L’-band vector vortex coronagraph installed  in the near-infrared  camera and spectrograph  NIRC2 at the Keck II telescope, with the aim of unveiling the nature of the spiral structure by constraining the presence of planetary companions in the system.}
   {Our high-contrast imaging observations show a bright ($\Delta L'=7.0\pm 0.3$ mag) point-like emission, south of MWC~758 at a deprojected separation of $\sim$20~au ($r=0\farcs 111 \pm 0\farcs 004$) from the central star. We also recover the two spiral arms (south-east and north-west), already imaged by previous studies in polarized light, and discover a third one to the south-west of the star. No additional companions were detected in the system  down to 5 Jupiter masses beyond 0\farcs6 from the star.}
   {We propose that the bright L'-band emission could be caused by the presence of an embedded and accreting protoplanet, although the possibility of it being an asymmetric disk feature cannot be excluded. The spiral structure is probably not related to the protoplanet candidate, unless on an inclined and eccentric orbit, and it could be due to one (or more) yet undetected planetary companions at the edge of or outside the spiral pattern. 
   Future observations and  additional simulations will be needed to shed light on the true nature of the point-like source and its link with the spiral arms.}
   {}


\maketitle


\section{Introduction} \label{sec:intro}

Understanding how planet formation takes place is a fundamental question in astronomy today. The $\sim$3000 planets discovered in the last two decades thanks to various techniques allow astronomers to study planet mass and orbital parameter distributions, and their dependence on the properties of the host stars.
Despite the important advancements made in the knowledge of planetary systems, many aspects concerning the initial conditions for planet formation and evolution still remain unknown.
High angular resolution imaging of young protoplanetary disks in the closest star forming regions could provide answers to these questions. A variety of disk structures have been detected through infrared scattered light or mm-wave imaging. Circumstellar disks may present large cavities \citep[e.g.,][]{andrews2011}, gaps and bright rings \citep[e.g.,][]{Quanz2013b}, asymmetries \citep[e.g.,][]{vanderMarel2013}, and spiral arms \citep[e.g.,][]{Garufi2013,benisty2015}. In some cases the combination of polarized scattered light and millimeter measurements has even shown spatial segregation, which could be directly linked to the presence of planets \citep{pinilla2012}.
Direct images of young planets embedded in protoplanetary disks would offer the possibility of investigating the link between the initial stages of planet formation and the final outcomes of the process.

In this context, MWC~758 (HD~36112) offers a unique environment to probe the existence of planetary companions and to explore  the connection between disk structures and planet formation.
MWC~758  is a young stellar object \citep[$3.5 \pm 2$~Myr,][]{Meeus2012} at a distance of 151$^{+9}_{-8}$ pc \citep{Gaia2016} close to the edge of the Taurus star-forming region (stellar properties are given in Table~\ref{tab:stellar properties}). 
Measurements of resolved CO emission around the star determined the stellar mass to be $2.0 \pm 0.2 M_{\odot}$ and the disk to have an inclination of $21\degr \pm 2\degr$ and a position angle of the semi-major axis of $65\degr \pm 7\degr$ \citep{Isella2010}. The mass and age estimates were based on the previously adopted Hipparcos distances of 200~pc \citep{vandenAncker1998} and 279~pc \citep{vanLeeuwen2007}. 
Given the revised Gaia distance, the star could be older and lighter than thought so far.  In this paper, we assume a stellar mass of $1.5 \pm 0.2 M_{\odot}$, reflecting the  scaling of the dynamical mass estimate to the new Gaia distance.
Based on its SED, MWC~758 has been classified as pre-transition disk \citep{grady2013}. Although a cavity of $55$ astronomical units (au) in radius has been inferred from dust millimeter emission \citep{andrews2011}, infrared polarized intensity observations found no clear evidence for a cavity in scattered light \citep{grady2013,benisty2015}. Using Ks-band (2.15~$\mu$m) direct imaging and H-band (1.65~$\mu$m) polarimetric imaging with the High Contrast Instrument with Adaptive Optics (HiCIAO) at the Subaru Telescope, \citet{grady2013} detected two spiral arms and polarized light down to 0\farcs1 (15~au) from the star. Recent VLT Spectro-Polarimetric High-contrast Exoplanet REsearch (SPHERE) observations in the Y band (1.04 $\mu$m) have confirmed the presence of scattered light at least down to 14~au \citep{benisty2015}.
The asymmetries observed by \citet{Isella2010} in the mm-dust distribution and in CO emission suggest that the disk may be gravitationally perturbed by a low mass companion orbiting within a radius of 23~au (assuming a distance of 151~pc). The asymmetric cm-dust distribution was shown to follow the location of the mm-dust \citep{Marino2015}, hinting towards the hypothesis of a dust trap, which could also be created by such companion in the gap through the Rossby wave instability \citep[e.g.,][]{Pinilla2012b}. 
Hydro-dynamical simulations of the disk indicate that the observed spirals could instead be launched by a massive planet or brown dwarf at larger separations \citep[$\sim$100~au based on the revised Gaia distance,][]{Dong2015b}.
The presence of stellar companions down to a mass limit of 12 $M_{\mathrm{Jup}}$ at 0\farcs25 and of planets outside 0\farcs5 \citep[5 $M_{\mathrm{Jup}}$ at 0\farcs5, and 3 $M_{\mathrm{Jup}}$ at 1\arcsec, according to the BT-SETTL models;][]{allard2012} has been ruled out based on a combination of sparse aperture masking observations at L' band and angular differential imaging at K' and Ks bands \citep{grady2013}. 

In this paper we present high contrast imaging observations of the Herbig Ae star MWC~758 obtained with the NIRC2 camera at the Keck II telescope in the L' band (3.8 $\mu$m).
Thanks to the use of the Keck adaptive optics system, combined with the recently commissioned vortex coronagraph and with high-contrast differential imaging techniques, the observations of MWC~758 presented in this letter achieved unprecedented sensitivity in the innermost 0\farcs25 and allowed us to probe the existence of planetary companions down to 0\farcs08.
In Sect.~\ref{sec:observations}, we summarize the observations that we carried out and the data reduction process. The results are described in Sect.~\ref{sec:results}. We discuss the nature of the point like source and the origin of the spiral arms in Sects.~\ref{sec:discussion - bright emission} and~\ref{sec:spiral arms}, respectively. Finally in Sect.~\ref{sec:conclusions} we present our conclusions.

\begin{table}[t]
\caption{Stellar Properties}             
\label{tab:stellar properties}      
\centering                                      
\begin{tabular}{c c}          
\hline\hline                        
Properties & Values \\    
\hline                                   
RA (J2000)  & 05$^{\rm h}$30$^{\rm m}$27$\fs$530 \\
DEC (J2000)  & +25\degr19\arcsec 57\farcs082  \\
Age (Myr) & $3.5\pm2.0 \; ^{(1)}$\\
Mass (M$_{\odot}$) & 1.5 $\pm$ 0.2 \; $^{(2)}$\\
L' (mag) & 4.75 $^{(3)}$\\
Distance (pc) & ${151}^{+8}_{-9}$ $^{(4)}$\\
\hline                                             
\end{tabular}
\tablebib{
(1)~\cite{Meeus2012}; (2) we scaled the dynamical mass estimate in \citet{Isella2010} to a distance of 151~pc; (3) \citet{Malfait1998}; (4) \citet{Gaia2016}.
}
\end{table}


\section{Observations and Data Reduction} \label{sec:observations}

MWC~758 was observed twice (see Table~\ref{tab:observations}) with the Keck II telescope at W.~M.~Keck Observatory, taking advantage of the L’-band vector vortex coronagraph installed in the near-infrared camera and spectrograph NIRC2 \citep{serabyn2017}. This vector vortex coronagraph is a phase-mask coronagraph enabling high contrast imaging close to the diffraction limit of the telescope ($\sim$0\farcs08). On October 24, 2015 we obtained 33 minutes (80 frames) of on-source integration time and 129\degr\ of field rotation (see Table~\ref{tab:observations}) to allow for angular differential imaging \citep[ADI,][]{marois2006}.  Each frame is the sum of 50 internally co-added frames of 0.5~s discrete integration time (DIT) each. During the coronagraphic acquisitions, the sky ($\text{DIT}=0.5$~s) and the unsaturated stellar point-spread function (PSF, $\text{DIT}=0.018$~s) were also measured for background subtraction and photometric calibration purposes, respectively. 
The alignment of the star onto the coronagraph center is crucial for high contrast at small angles. In this case it was performed using the tip-tilt retrieval technique QACITS \citep{huby2015,huby2017}, as already implemented for HD~141569 \citep{mawet2017}.  Thanks to QACITS, we could reach a centering accuracy of $0.03\lambda/D$~rms (3~mas rms). 
MWC~758 was then re-observed on October 24, 2016, following the same observing strategy as for the first epoch of observations (detailed information are given in Table~\ref{tab:observations}). Due to cirrus clouds the weather conditions during this run were not as good as in the 2015 run. 

\begin{table}[t]
\caption{Observations}              
\label{tab:observations}     
\centering                                     
\begin{tabular}{c c c}         
\hline\hline                        
  & First epoch & Second epoch \\    
\hline                                   
UT date (yyyy/mm/dd) & 2015/10/24&  2016/10/24\\  
DIT (s)  & 0.5 & 0.25 \\
Coadds  & 50 & 160 \\
Number of frames & 80 & 80\\
Total Int. Time (s) & 2000 & 3200 \\
Plate scale (mas/pix) & 9.942 & 9.942 \\
Filter Coronagraph & L' &  L' \\
Par. angle start/end ($^{\circ}$) & -128/+103 &  -90/+97\\
Mean airmass & 1.012 & 1.074 \\
Median seeing (\arcsec) & 0.64 & 0.75 \\
\hline                                             
\end{tabular}
\end{table}

Both data sets were preprocessed using the Vortex Image Processing package \citep[VIP,][]{gomez2017}. Images were divided by a flat field obtained without the vortex phase mask and the background emission (sky) was subtracted based on principal component analysis \citep[PCA,][]{gomez2017,hunziker2017}. Bad pixel correction and bad frame removal were also applied to the data. Finally, the frames were re-centered through Fourier shift operations.
For each epoch, the stellar PSF was subtracted by performing PCA \citep{AmaraQuanz2012,Soummer2012} to the full set of frames.


\section{Results} \label{sec:results}

\begin{figure*}
   \centering
\includegraphics[width=15cm]{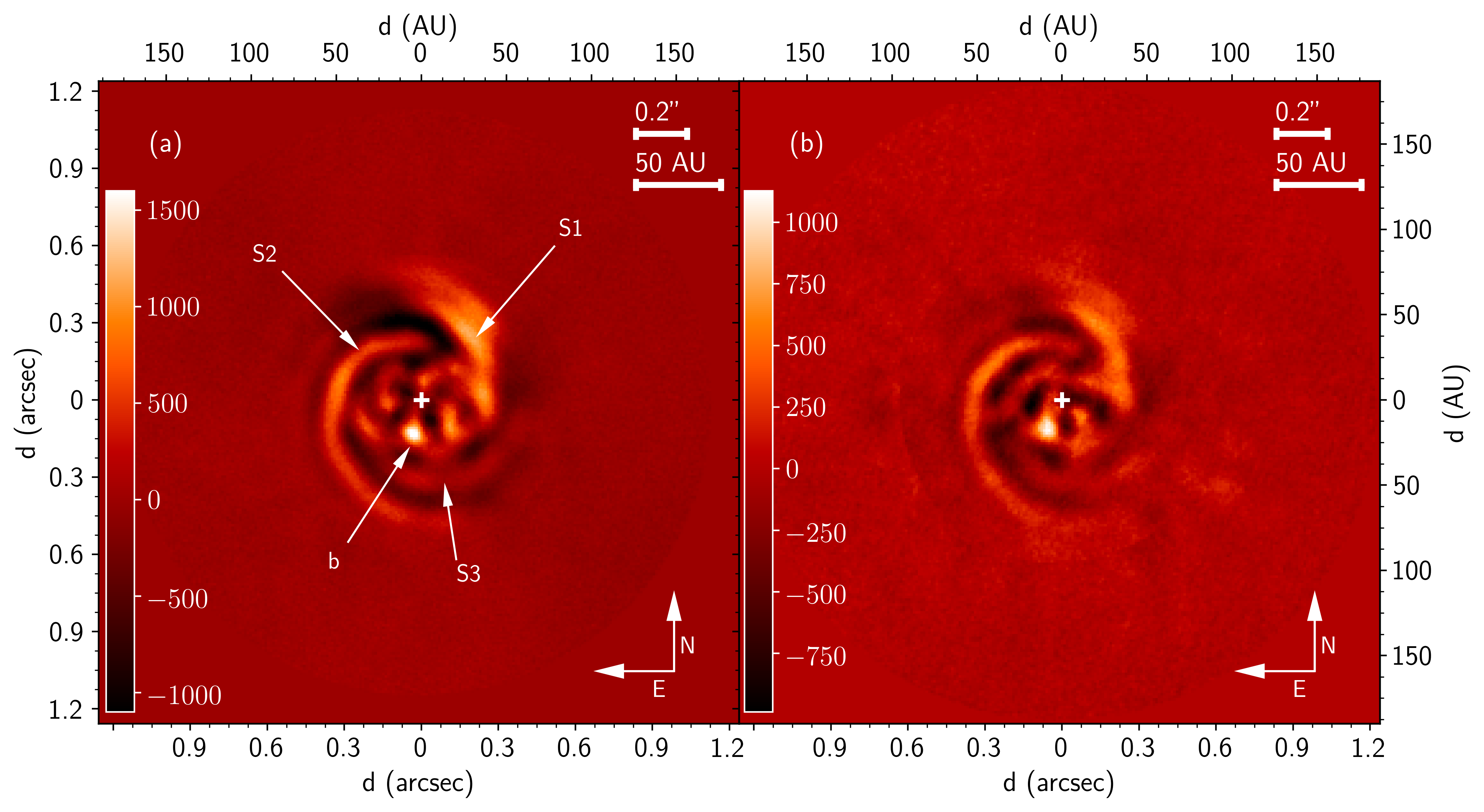}
      \caption{Final PCA-ADI images for the 2015 (a) and 2016 (b) data sets. Three spiral arms and a bright point-like feature are detected in the images. The three spiral arms and the point-like source are labeled with S1, S2, S3, and $b$, respectively.}
         \label{Figure:emission source}
   \end{figure*}

The final PCA-ADI L'-band images (Figure~\ref{Figure:emission source}) show a bright point-like emission source, detected south of MWC~758 (labeled $b$ in Figure~\ref{Figure:emission source}a), at $\sim$0\farcs1~from the central star. The images also recover the two spiral arms (S1 and S2) already observed in near-infrared polarized light \citep{grady2013,benisty2015}, and reveal an additional one to the SW (S3). 

In the following sections (\ref{subsec:bright emission}, \ref{subsec:upper limits}, and in \ref{subsec:spirals}) we present the results in details.

\subsection{The point-like source in the disk} \label{subsec:bright emission}

\begin{figure}
   \centering
\includegraphics[width=9cm]{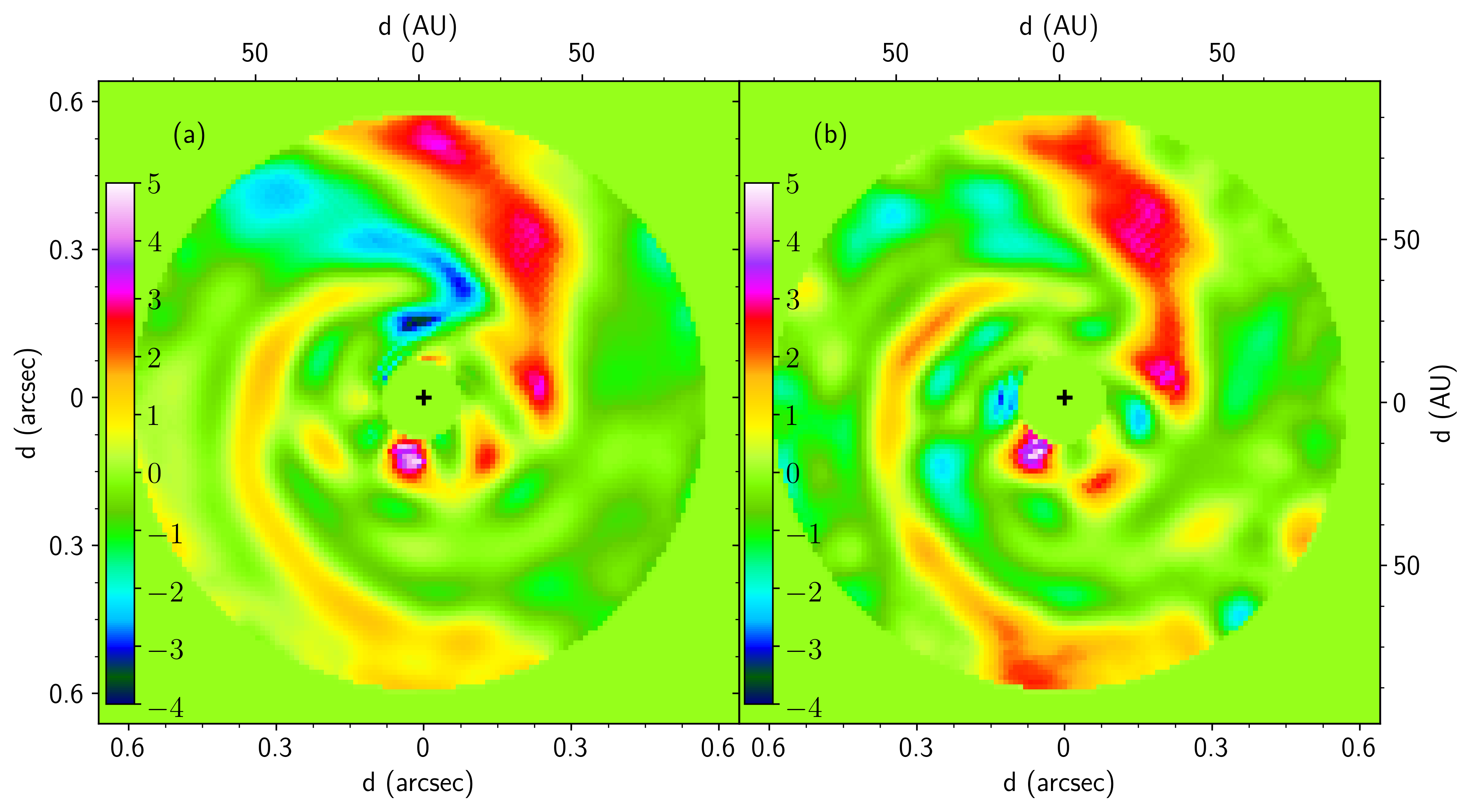}
      \caption{S/N maps for the 2015 (a) and 2016 (b) data sets. Apart from $b$, none of the other bright features in the inner part of the disk (inside 0\farcs2, down to the inner working angle at 0\farcs08), is recovered with a S/N $>$ 3}
         \label{Figure:snr maps}
   \end{figure}

In the PCA-ADI residual images for both epochs, a bright L'-band emission source is located at the same position interior to the spiral arms (see Figure~\ref{Figure:emission source}). For each dataset, the final image corresponds to the number of principal components that maximize the signal to noise ratio (S/N) of the point-like feature (3 and 9 components, respectively). To compute the S/N, we follow the \citet{mawet2014} prescription, where the signal is summed in a 1 full width at half maximum (FWHM) aperture around a given pixel, and the noise is computed as the standard deviation of the fluxes inside 1 FWHM apertures covering a 1 FWHM-wide annulus at the same radial distance from the center of the frame, taking into account the small sample statistic correction.
In both data sets, the point-source is recovered with a S/N of $\sim$5.  Figures~\ref{Figure:snr maps}a and \ref{Figure:snr maps}b show the S/N maps for the 2015 and 2016 final PCA-ADI images, respectively. None of the other bright features in the inner part of the disk (within 0\farcs2), is recovered with a S/N $>$ 3.

To assess the reliability of the detection, we performed a series of tests.
We both varied the number of PCA coefficients and divided the two data sets into subsets containing either half or a third of the frames, but covering the full field rotation.
In both cases, the emission source south of MWC~758 appears to be the most significant feature in the final PCA-ADI images. We also inverted the parallactic angles of the frames, but we could not artificially generate any feature as bright as the one we detected.

The astrometry and photometry of the source are determined by inserting negative artificial planets in the individual frames, varying at the same time their brightness and location. The artificial companions are obtained from the unsaturated PSF of the star, which was measured without the coronagraph. The brightness and position that minimize the residual in the final images are estimated through a standard Nelder-Mead minimization algorithm.
In the first epoch, the source is located at a distance $r = 0\farcs112 \pm 0\farcs006$ from the central star at a position angle $\text{PA}= 169\degr \pm 4\degr$, with a magnitude difference $\Delta L = 7.1 \pm 0.3$~mag. In the second dataset, the estimated position is $r =0 \farcs 110 \pm 0\farcs 006$ and $\text{PA} = 162\degr \pm 5\degr$, and the flux ratio is $\Delta L' = 6.9 \pm 0.5$~mag.
The magnitude difference takes into account the vortex transmission ($\sim$50\%) at these separations.
The uncertainties on the quantities due to speckle noise were determined by injecting a series of fake companions around the star in the raw, companion-subtracted cube at the same radial distance, and calculating the median errors of the retrieved  distributions. Variations of the total flux in the unsaturated PSF during the observing sequence were also included in the uncertainty on the brightness difference.
The two measurements of separation, position angle and contrast are consistent with each other within $1 \sigma$.
If the spirals are trailing, a companion in the disk is expected to rotate clock-wise. Given the distance of the source from the star and the time difference between the two epochs (1 year), orbital motion on a circular orbit would produce a displacement of $\sim$5\degr, which is in line with our measurements.

\subsection{Upper limits on other companions} \label{subsec:upper limits}

Apart from $b$, the only other point-like source observed in the field of view is the star located at 2\farcs3 to the NW, identified by \citep{grady2013} as background source.
In order to calculate robust detection limits, assess the mass constraints for other companions around MWC 758, and mitigate some of the shortcomings of standard contrast curves, we adopted the concepts of false positive fraction ($\mathrm{FPF}$) and true positive fraction \citep[$\mathrm{TPF}$,][]{wahhaj2013,ruane2017}. 

\begin{figure}[t]
\centering
\resizebox{\hsize}{!}{
\includegraphics[height=1.6cm]{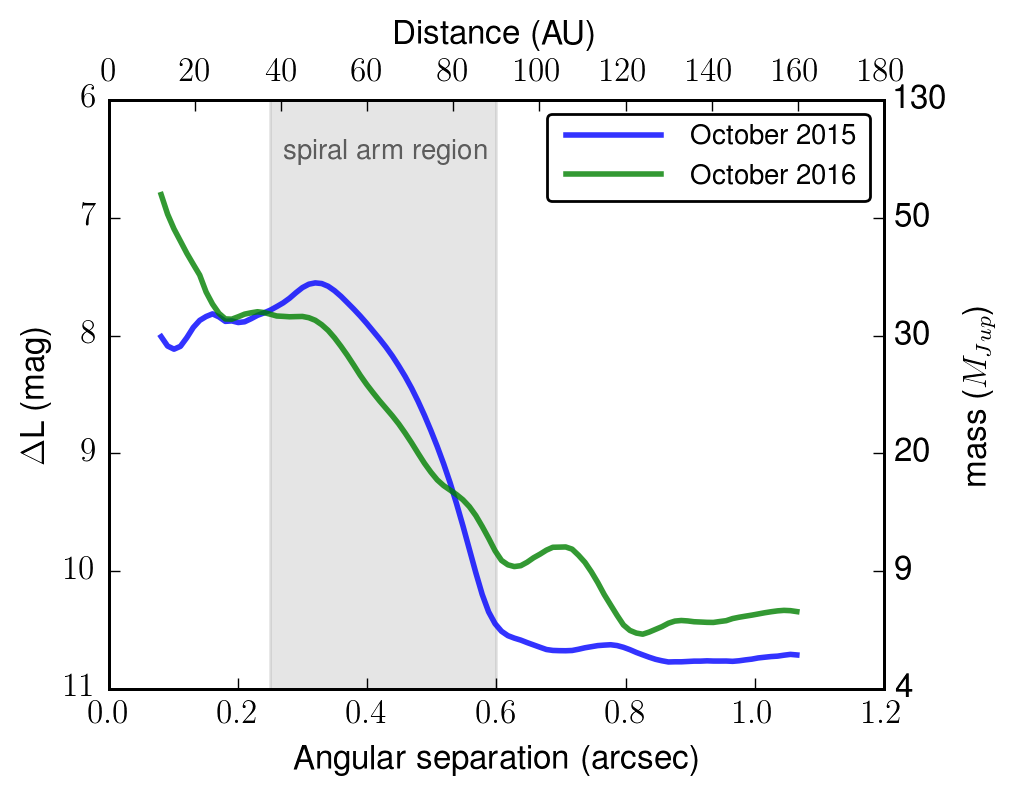}}
  \caption{Detection limits at 95\% completeness around MWC~758 for the Oct. 2015 (\emph{blue line}) and the Oct. 2016 (\emph{green line}) data sets. A false alarm probability of 1\% is allowed in our field of view (see Sect.~\ref{subsec:upper limits}). The inner region indicated by the gray area is dominated by the presence of the spirals arms. Outward, our detection limits for the Oct. 2015 data set are close to 5  $M_{\mathrm{Jup}}$, according to the BT-SETTL evolutionary model \citep{allard2012}. }
     \label{Figure:contrast curve}
\end{figure}

In the hypothesis that after subtracting the stellar PSF the noise in the image can be considered Gaussian \citep{mawet2014}, and that we set an acceptable fraction (1\%) of false positive in our field of view (a region of 1\arcsec $\times$ 1\arcsec), we computed the signal level corresponding to a given completeness level \citep{ruane2017}.
We adopted a 95\% completeness, or $\mathrm{TPF}=0.95$, which means that such a signal would be detected 95\% of the time.
The contrast curve at 95\% completeness is plotted as a function of angular distance in Figure~\ref{Figure:contrast curve}.
In the inner 0\farcs6, the sensitivity is limited by the scattered light emission from the spirals between 30 and 90~au.
Beyond 0\farcs6, the sensitivity becomes almost constant (background-limited) and we can exclude planetary-mass companions down to $\sim$5 M$_{\mathrm{Jup}}$ (see Figure~\ref{Figure:contrast curve}) according to the BT-SETTL evolutionary model \citep{allard2012}.

\subsection{Spiral arms} \label{subsec:spirals}

\begin{figure}[t]
\centering
\includegraphics[height=7cm]{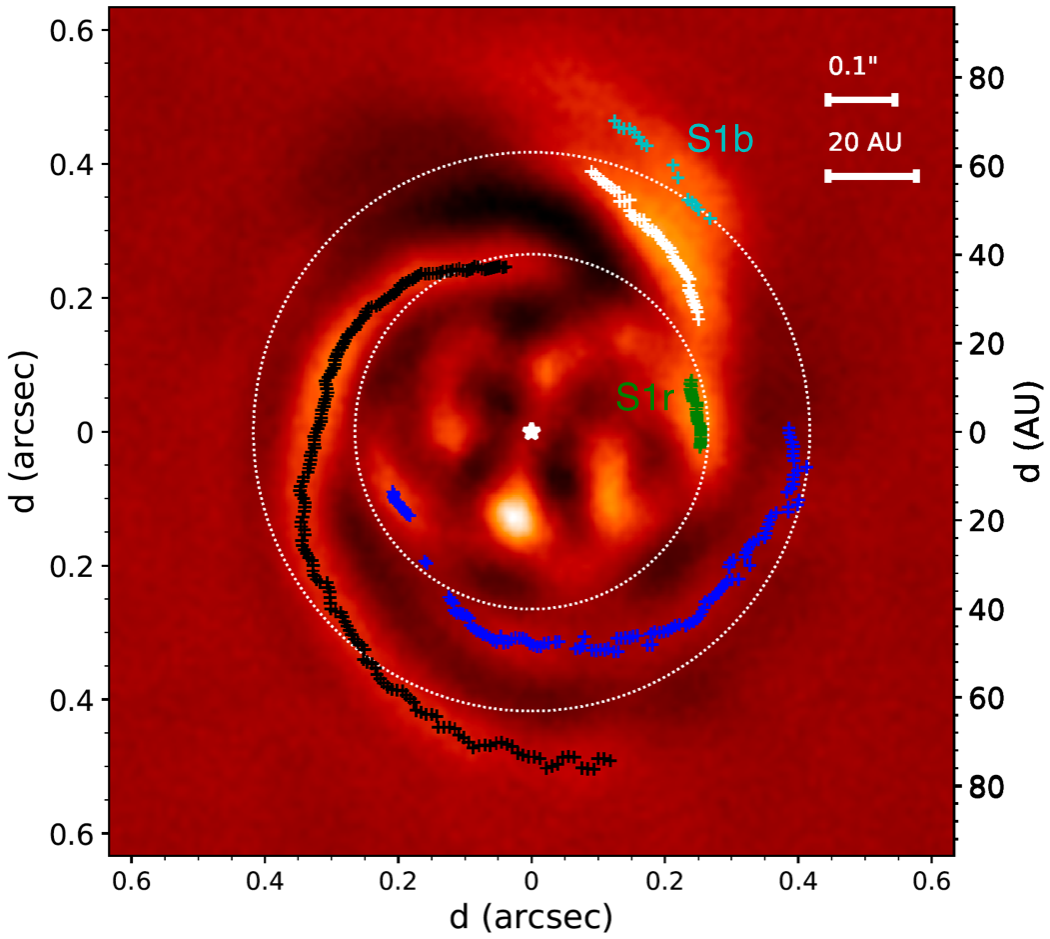}
  \caption{Deprojected disk image from the 2015 data set showing the trace of S1 (\emph{white}), S2 (\emph{black}), S3 (\emph{blue}), features \emph{i} (\emph{green}) and \emph{ii} (\emph{cyan}).  The dotted circles have 40~au and 63~au radial separations, and represent the limits within which the separation angle is computed between each pair of spirals (Figure~\ref{Figure:sep_angle}).}
     \label{Figure:spiral traces}
\end{figure}

Both the Oct.~2015 and Oct.~2016 final images (Figure~\ref{Figure:emission source}a and b) show the two bright spiral arms (S1 and S2) already detected in H- and Y-band polarized light \citep{grady2013,benisty2015}, and a third fainter spiral to the SW (S3), which has not yet been reported. A detailed discussion on the reliability of the detection of the third arm is given in Appendix~\ref{App:reliability spirals}. 
In addition to the three spirals, we recover the bright second arc, shifted radially outward from S1, observed in polarized light at $\text{PA} \sim 325\degr$ by \citet{benisty2015}. We refer to it as feature \emph{ii} in Figure \ref{Figure:spiral traces}.

Comparison of our image with the polarized light images of \citet{benisty2015} also reveals several differences. First, the bright Y-band arc located at $\sim$0\farcs2 separation and covering a PA range of 180\degr-- 270\degr \citep[feature (2) in][]{benisty2015} appears much less prominent in L'-band,  
but is still recovered at the same location as in the polarized light image. A possible reason for the significant damping of the arc in our image is the tendency of ADI to self-subtract extended axi-symmetrical signal. 
Second, while S1 and the Y-band arc appear smoothly connected in the polarized image, this is not the case in our L'-band images, where a clumpy structure is distinguished close to the root of S1 (feature \emph{i} in Figure \ref{Figure:spiral traces}). Part of the reason for such discontinuity could be the ADI filtering of a slightly over-illuminated area of the spiral, as ADI is known to produce negative azimuthal lobes. Alternatively, this could indicate that it is tracing a different physical process (e.g. the merging of S1 and S2).

To characterize the spiral arms and features \emph{i} and \emph{ii}, we identified their trace as radial intensity maxima in azimuthal steps of 1\degr. This was done both in the final PCA-ADI image and in a deprojected one, based on a disk inclination of 21\degr~and a PA of the semi-major axis of 65\degr~\citep{Isella2010}. The deprojected image is shown in Figure~\ref{Figure:spiral traces}. Only the Oct.~2015 PCA-ADI final image was used, as it reaches better sensitivity than the Oct.~2016 data set.
The trace of the spirals then allowed us to measure their pitch angles, defined at each point as the angle between the tangent to the spiral and the local azimuthal vector. For each trace, the pitch angle was estimated in two different ways: (a) we considered the average value of the pitch angle computed from all pairs of consecutive points in the trace, and (b) we derived the pitch angle of the best fit logarithmic spiral given by $r=a e^{b\theta}$. Logarithmic spiral arms have the property of keeping a constant pitch angle throughout their length, given by the complementary of $\arctan(1/b)$. Both methods yielded consistent measurements, although method (a) led to larger uncertainties.

\begin{figure}[t]
\centering
\includegraphics[width=8cm]{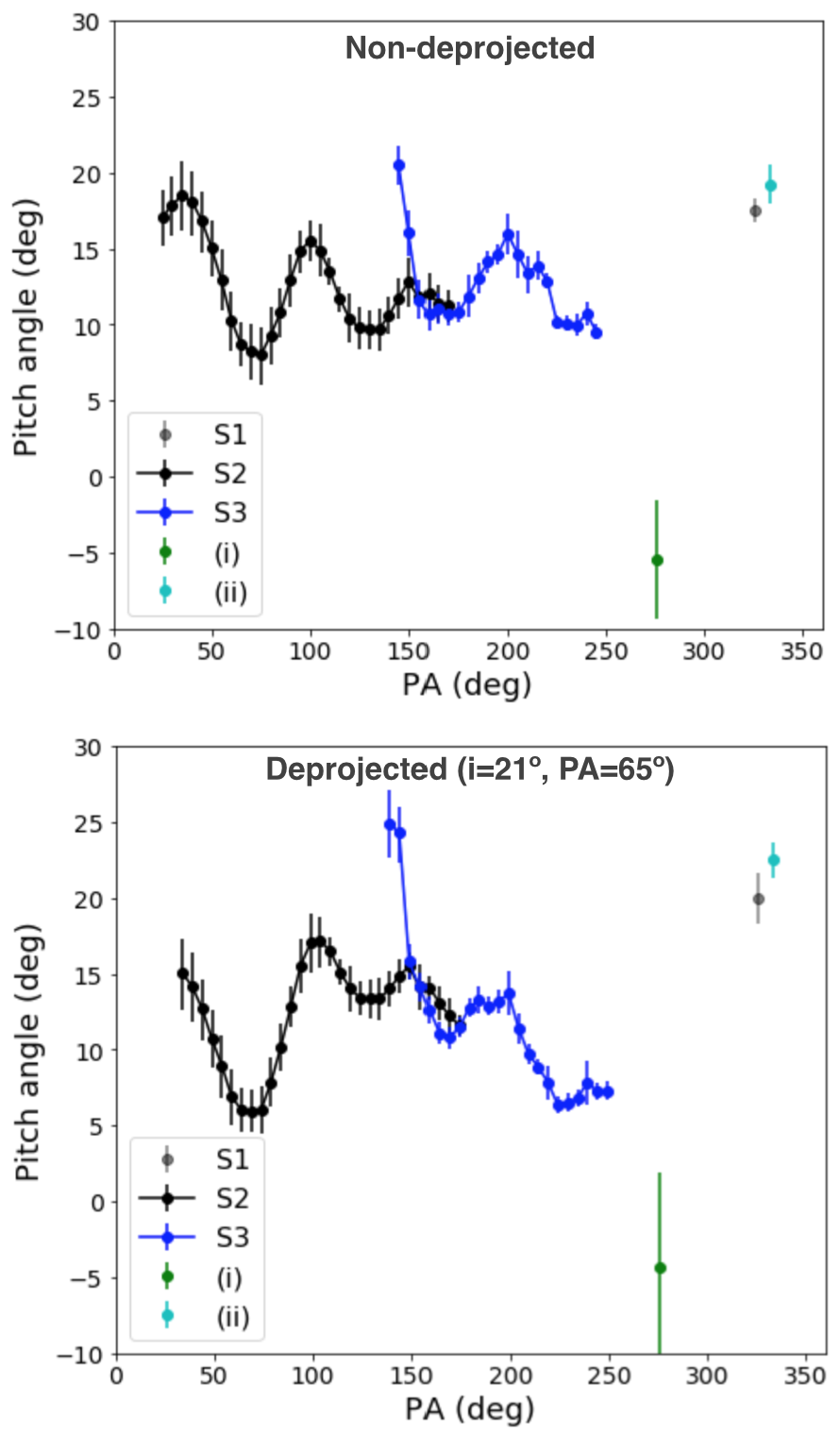}
  \caption{Pitch angle of each spiral, in the non-deprojected (top) and deprojected image (bottom). For each measurement, we considered either the whole trace (for S1, features \emph{i} and \emph{ii}), or  consecutive arcs subtending 50\degr~(for S2 and S3) to trace the evolution of the pitch angle along the trace in order to provide reliable pitch angle estimates. The estimate for feature \emph{ii} has a larger uncertainty because its trace only subtends $\sim$25\degr, instead of $\sim$50\degr for S1 and feature \emph{ii}.}
     \label{Figure:pitch angle}
\end{figure}

Figure \ref{Figure:pitch angle} shows the pitch angles measured with method (b) for each feature identified in Figure~\ref{Figure:spiral traces}. 
We measured the evolution of the pitch angle over S2 and S3 using method (b) on consecutive sections of the spirals separated by 5\degr~and subtending each 50\degr. A single measurement using method (b) is provided for S1, features \emph{i} and \emph{ii}, due to the short PA range they subtend.
We first notice that the pitch angle of  feature \emph{i} is slightly negative (quasi-null), which suggests that it is more likely related to the Y-band circular arc rather than to S1.
We also note that the root of S3 (PA $<$150\degr) shows a more significant pitch angle than the rest of S3. This could imply that it is either tracing a feature of the disk unrelated to S3, or that ADI filtering significantly alters the shape of S3 at such close separation. Overall, S1 and feature \emph{ii} present a slightly larger pitch angle than both S2 or S3. The difference is more significant after deprojection ($\sim$20\degr~instead of $\sim$7--16\degr). It appears thus more likely that feature \emph{ii} is related to S1 rather than S3, although the possibility that feature \emph{ii} traces the outer tip of S3 cannot be completely ruled out in view of the strong fluctuations in the pitch angle of both S2 and S3 (up to 10\degr). The distortion of the spirals induced by the disk deprojection also appears to enhance the drop in pitch angle along the PA of the semi-major axis ($65\pm7$\degr and $245\pm7$\degr) for both S2 and S3. Implications are further discussed in Sec.~\ref{sec:spiral arms}.


\section{Discussion} \label{sec:discussion}
\subsection{The nature of the point-like source} \label{sec:discussion - bright emission}

If we take a mean weighted by the errors of the estimates of the two epochs, the point-like source is located at an angular separation of 
$0\farcs 111 \pm 0\farcs 004$ with a magnitude difference of $\Delta L'=7.0 \pm 0.3$~mag.
Given the distance of the star \citep[151 pc;][]{Gaia2016}, and assuming the disk to be nearly face on \citep{Isella2010}, the physical separation from the central star is $16.7 \pm 0.6$~au. Taking into account the inclination of the disk, the bright emission would be at 20$\pm$1~au (assuming
co-planarity, as shown in Figure \ref{Figure:spiral traces}). 

\begin{figure}
   \centering
   \resizebox{\hsize}{!}{
   \includegraphics[height=1.4cm]{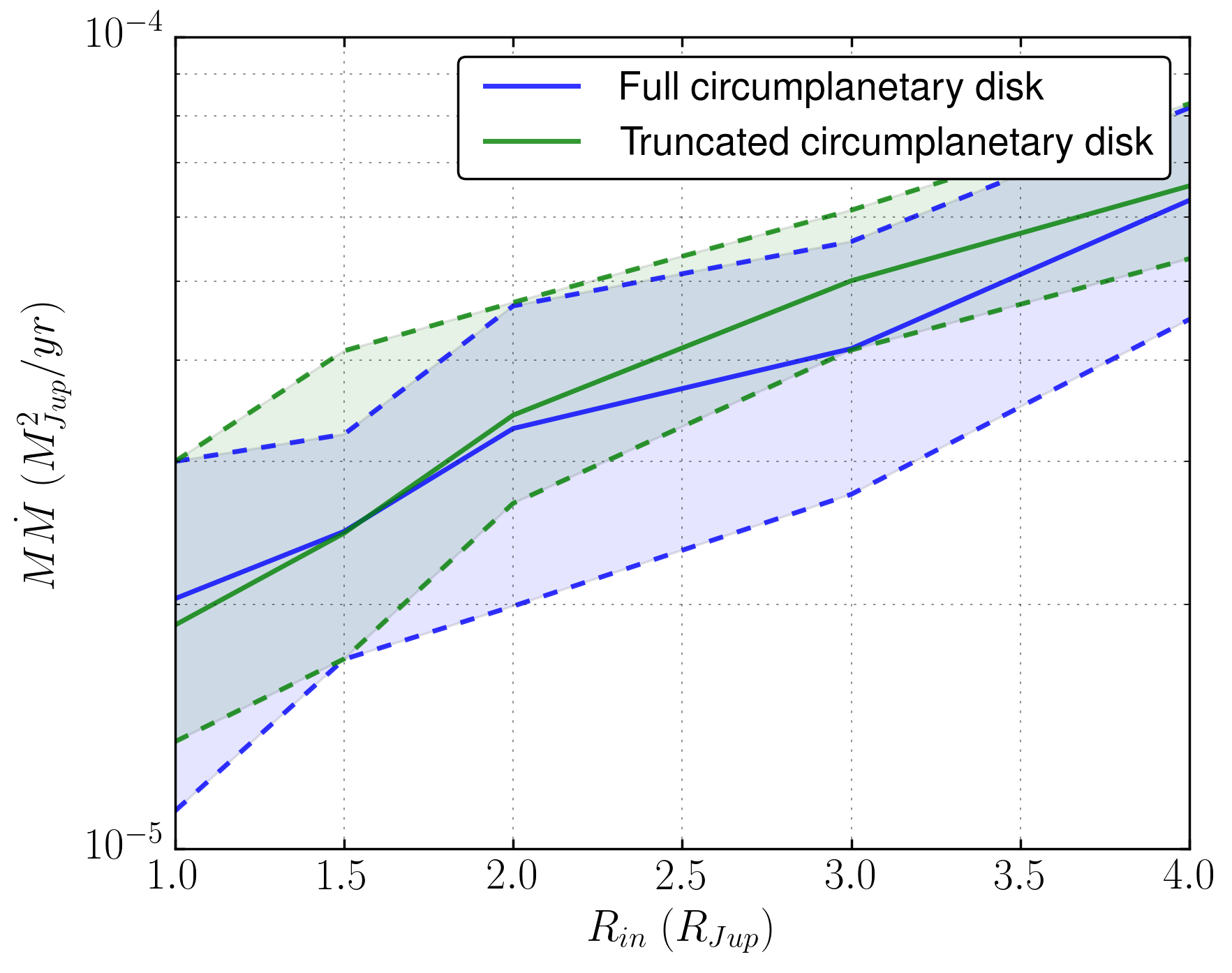}}
      \caption{Circumplanetary disk accretion models from \cite{zhu2015}. The product of the mass of the planet and the disk accretion rate ($M_{p}\dot{M}$) changes as a function of the disk inner radius ($R_{in}$). The solid blue and green lines represent the $M_{p}\dot{M}$ vs. $R_{in}$ curve for the measured L'-band absolute magnitude ($M_{L'}=5.9$~mag), in the case of a full or a truncated circumplanetary disk, respectively. The dotted lines indicate the 1$\sigma$ error bars.}
         \label{Figure:accretion models}
   \end{figure}

To explain the nature of the point-like source, we explored different possibilities. Comparing the separation and the brightness of the source with the TRILEGAL tridimensional model of the galaxy \citep{girardi2012}, we can safely reject the hypothesis that the bright emission is a background star (probability $\simeq 10^{-6}$). If the L'-band emission came from the photosphere of a low-mass companion, its mass would range between 41-64 Jupiter masses ($M_{\mathrm{Jup}}$), according to the BT-SETTL evolutionary models \citep{allard2012}. However, the non-detection of a fully depleted cavity in micrometer-size dust, requiring a steady replenishment of small particles, restricts the mass of companions in the inner disk of MWC~758 to be $\lesssim$ 5.5 $M_{\mathrm{Jup}}$ \citep{pinilla2015}, or even smaller depending on the assumed model \citep[e.g. a $1M_{\mathrm{Jup}}$ planet should already start opening a gap in the gas distribution, ][]{dong2017b,fouchet2010,paardekooper2004}. With such mass constraints, only a protoplanet (hence MWC~758~b) surrounded by a circumplanetary accretion disk could account for the observed brightness. According to the circumplanetary disk accretion models of \citet{zhu2015}, its L'-band luminosity is compatible with a 0.5-5~$M_{\mathrm{Jup}}$ planet accreting at a rate of 10$^{-7}$-10$^{-9}$ $M_{\odot}$ yr$^{-1}$ (see Figure~\ref{Figure:accretion models}). The corresponding K-band magnitude difference with the star ($\sim$7--9 mag) would be consistent with the non-detection of MWC~758~b in previous images at this wavelength \citep{grady2013}. Due to the highly structured nature of the inner disk, we cannot discard the possibility that the point-like source is associated with an asymmetric disk feature, as recently suggested for HD~169142 \citep{Ligi2017}. However, only future observations, as for instance the lack of orbital motion or the detection (or not) of H$\alpha$ emission, can test this hypothesis.

Besides these potential astrophysical origins, we also consider the possibility that the point-like source corresponds to a false positive detection, i.e., a bright speckle appearing twice at the same location. We show in Appendix~\ref{App:Odds Ratio} that, under Gaussian noise assumption, the astrophysical explanation is favoured at an odds ratio of $\sim$1000:1 with respect to the false positive hypothesis.

\subsection{The spiral arms} \label{sec:spiral arms}

Three spiral arms are detected at L'-band, which most likely traces scattered stellar light by sub-micrometer size dust. The detection of S3 in L'-band and its non-detection with polarized light at shorter wavelength could be explained by  
the different dust scattering properties at L'-band (3.8 $\mu$m) compared to Y-band (1.04 $\mu$m). If the emission traces Rayleigh scattering of sub-micrometer size grains (cross-section $\propto \lambda^{-4}$), then Y-band would trace mostly the disk surface, while L'-band could probe deeper layers of the disk. Therefore, the non-detection of S3 at Y-band could mean that it has a smaller scale height than S1 and S2. The different appearance of feature \emph{i} at Y- and L'-band could also be due to the tracing of different disk layers, or to ADI filtering, as discussed in Sec.~\ref{subsec:spirals}. 
Part of the L'-band emission could also trace shocks occurring in the spiral wake, as they could significantly heat the disk locally \citep[e.g.,][]{Richert2015,Lyra2016}. However, \citet{Rafikov2016} argues that the increase in temperature due to shocks should be negligible, meaning that the observed spirals would only trace scattered stellar light.

Regarding the origin of feature \emph{ii}, we note a striking resemblance with the double arc seen in the disk of HD~100453 \citep{Benisty2017}. It was suggested that such feature can be explained as the scattering surface 
of the bottom side of the disk. We note that in the case of MWC~758 this explanation is consistent with the estimated inclination and PA of the outer disk ($i \sim 21$\degr, PA $\sim 65$\degr), since such double feature is expected to be more prominent along the semi-minor axis of the disk. If this interpretation is correct, the NW side of the outer disk is closer to us than the SE side.

Could any of the observed spirals be launched by the candidate companion?
Hydro-dynamical simulations and corresponding radiative transfer models suggest that observable spirals in near-infrared light could only be launched by companions massive enough for the linear spiral density wave theory to break down \citep{juhasz2015,dong2017}. In this case, the planet responsible for the spiral structure would be located at the edge of or outside the spiral arms in order to be able to reproduce the observed pitch angles \citep{Dong2015b}, indicating that the protoplanet candidate is most likely not responsible for the observed disk structure. 
However, the possibility that the candidate companion is on an inclined, eccentric orbit cannot be ruled out. Recent simulations suggest that a mild perturbation on the inclination of a companion in the inner disk can lead to a polar orbit within a few Myr \citep{Owen2017,MartinLubow2017}. In such case, spiral predictions from circular orbits co-planar with the disk would not be valid.
Recent simulations in the case of the disk of HD~142527 indicate that the close-in companion with an inclined and eccentric orbit \citep{Lacour2016} is able to qualitatively reproduce the spiral arm pattern in the outer disk (Price et al. 2017, in prep.).
Similarly, some or all of the observed spirals in MWC~758 could be launched by the candidate companion if in the same plane as the inner disk, which is  likely misaligned with respect to the outer disk. The difference in inclination is estimated between 10\degr and 30\degr, while the PA does not appear well constrained \citep{Isella2006,Isella2008,lazareff2017}. 
Finally, the large fluctuations in the measured pitch angle along S2 and S3 could also be considered as a clue that the spirals are launched by a companion whose orbital plane is different than the plane of the outer disk. Indeed, if the spirals were seen face-on after deprojection, one would not expect significant variations in their opening angle \citep[apart in the direct vicinity of the companion,][]{zhu2015b,Juhasz2017}. Nevertheless, our deprojection does not take into account the flaring of the disk, which may bias the measured pitch angles.

\begin{figure}
\centering
\resizebox{\hsize}{!}{           
\includegraphics[height=0.6cm]{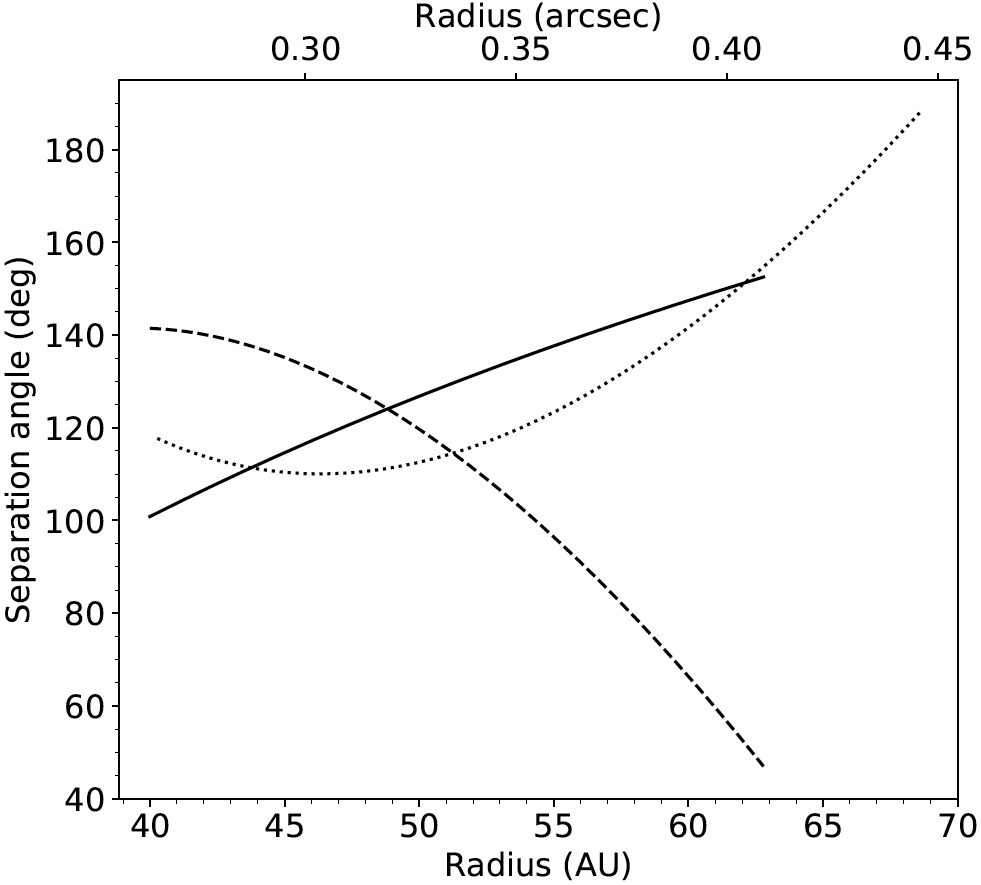}
}
\caption{
      Separation angle between S1 and S2 (\emph{solid line}), between S1 and S3 (\emph{dashed line}), and between S2 and S3 (\emph{dotted line}).}
    \label{Figure:sep_angle}
\end{figure}

In case a massive, yet undetected companion located outside of the spirals is needed to account for the observed multiple spiral pattern, an empirical relationship has recently been established between the mass of the companion and the separation angle $\phi_{\mathrm{sep}}$ between primary and secondary spiral arms: 
$\phi_{\mathrm{sep}}=102$\degr$ (q/0.001)^{0.2}$, where $q$ is the mass ratio between the companion and the star \citep[][hereafter FD15]{fung2015}. 
In order to investigate the origin of the spirals of MWC 758, we measured the separation angle $\phi_{\mathrm{sep}}$ between each pair of spiral arms between 40 and $\sim$63~au, where all spirals are clearly defined in our final PCA-ADI image (Figure~\ref{Figure:sep_angle}). These measurements are based on geometrical fits of the spirals, to be presented in Barraza et al. (2017, in prep.).
In the FD15 simulations, the separation angles are relatively constant with radius, with only a slight decreasing trend for all models with $\phi_{\mathrm{sep}} > 80$\degr.
Here, for each pair of spirals, the observed separation angle appears to vary significantly with radius. 
Only in the limited range of radii $\sim$40--55 au, $\phi_{\mathrm{sep}}$ for the S1-S3 and the S2-S3 pairs are comparable to the FD15 models for $q=4\times 10^{-3}$ and $q=10^{-3}$, respectively. Considering a stellar mass of $1.5M_{\odot}$, this would correspond to a companion mass of $\sim 6 M_{\mathrm{Jup}}$ (for S1-S3) and $\sim 2 M_{\mathrm{Jup}}$  (for S2-S3).
The first possibility is consistent with our detection limits inside 90~au, while the second is compatible with our sensitivity limits at any radius (Figure~\ref{Figure:contrast curve}).
Nevertheless, these predictions require the presence of a giant planet at large separation ($>80$~au), which is expected to be very rare \citep[e.g.][]{Vigan2017}.
Furthermore, in the case that S1-S3 (resp. S2-S3) are launched by an external companion, another companion would yet be required to account for S2 (resp. S1), except if S2 (resp. S1) happens to be a very bright tertiary arm, even brighter than the secondary arm (i.e.~S3 in both cases).

For the sake of completeness, we also consider the possibility that the observed spiral pattern is induced by one or several other physical mechanisms.
Hydro-dynamical simulations have suggested that shadows producing a periodical perturbation in temperature on the outer disk could be able to launch large scale symmetrical spiral arms \citep{montesinos2016}. In the case of an inclined inner disk with respect to the outer disk, two shadows are cast and a relatively symmetric two-arm spiral pattern forms in the outer disk.
Nevertheless, there is no detection of conspicuous shadows similar to the case of HD~142527 \citep{Avenhaus2014,Marino2015b} or HD~100453 \citep{Benisty2017}, which makes this possibility unlikely. 

The flyby by an external star is also known to produce a one- or two-arm spiral pattern \citep[e.g.,][]{Quillen2005}. We investigated the possibility that MWC 758 could have undergone a fly-by from the source located at 2\farcs3. As detailed in Appendix \ref{App:background source}, the color and magnitude of the object are not compatible with a young object at the distance of Taurus, and could either correspond to an early red dwarf at the distance of MWC~758, or a red giant star much farther away. The probability that an old red dwarf happens to cross the star forming region of Taurus is very small, we hence favor the second hypothesis (background red giant).

Based on mm-dust continuum observations and assuming a fiducial gas-to-dust ratio of 100, the disk mass was estimated to 0.008 $M_{\odot}$ \citep{andrews2011}. Considering a stellar mass of 1.5 $M_{\odot}$, the disk is about 50 times less massive than required to form spiral arms by self-gravity \citep{Dong2015c}.  We note that this is also an argument against the possibility of the spirals being launched by shadows, as that mechanism might require a marginally stable disk \citep{montesinos2016}.

\section{Conclusions} \label{sec:conclusions}

The L'-band vortex coronagraph installed on Keck/NIRC2 was used to discover a point-like source with $\sim$7.0 mag contrast in L'-band at only $\sim$0\farcs11 from the central star, and to reveal an additional spiral arm in the disk. 
Although the possibility of an asymmetric disk feature cannot be excluded, we argue that the $L' \sim 11.7$~mag apparent magnitude emission ($M_{L'} \sim 5.9$, assuming a distance of 151 pc) is most likely due to an embedded protoplanet. In this case the newborn planet would be sourrounded by an accreting circumplanetary disk, which may account for most of the observed luminosity.
The presence of scattered light down to 15~au \citep{benisty2015} indicates that the planet should be small enough to have only marginally affected the inner part of the disk.

We considered several possibilities for the origin of the spiral arms. Neither disk self-gravity nor the source at $\sim$2\farcs3 (most likely a background red giant) appear to be able to account for the observed structures. 
Our analysis suggests that the most likely explanation for the observed spiral pattern involves either (i) the candidate companion itself if it is on an inclined and eccentric orbit, possibly co-planar with the inner disk; (ii) the presence of an undetected $\sim 6 M_{\mathrm{Jup}}$ planet at the outer tip of S1; (iii) the presence of an undetected $\sim 2 M_{\mathrm{Jup}}$ planet at or outside the outer tip of S2; or a combination of the above explanations.
Both the second and third cases are compatible with our sensitivity constraints.
The second case is similar to the prediction based on previous hydro-dynamical simulations of the disk of MWC~758 \citep{Dong2015b}.

New dedicated simulations considering the revised distance of MWC~758 \citep[151 pc,][]{Gaia2016}, and including the protoplanet, the third spiral arm and our mass constraints in the outer disk, will enable to further constrain the origin of the observed spiral pattern. 
Future re-detection (or non-detection) of the companion at other wavelengths will help us constraining its properties thanks to the comparison with SED predictions \citep{zhu2015,eisner2015}.
Follow-ups with ALMA could probe the dynamics of the disk and also provide new insight on the nature of the bright L'-band point-source. If the observed feature is an accreting circumplanetary disk, it is expected to leave an observable kinematic signature in ALMA observations \citep{Perez2015}.
Its accretion nature could also be confirmed with direct imaging in the H$\alpha$ line, as for HD~142527B \citep{Close2014} or LkCa~15 \citep{sallum2015}.

\begin{acknowledgements}
We would like to acknowledge the anonymous referee for the helpful comments.
We thank Simon Casassus and Sebasti\'an Perez for fruitful discussions on the spiral arms, and Farzana Meru for useful insights on disk cavities. This research was founded by the European Research Council under the European Union's Seventh Framework Program 
(ERC Grant Agreement n. 337569) and by the French Community of Belgium through an ARC grant for Concerted Research Action. VC acknowledges support
from CONICYT through CONICYT-PCHA/Doctorado Nacional/2016-21161112, and from the Millennium Science 
Initiative (Chilean Ministry of Economy), through grant RC130007. GR is supported by an NSF Astronomy and Astrophysics Postdoctoral Fellowship under award AST-1602444. EC acknowledges support from NASA through Hubble Fellowship grant HST-HF2-51355 awarded by STScI, operated by AURA Inc. for NASA under contract NAS5-2655. This work used data reprocessed as part of the ALICE program, which was supported by NASA through grant HST-AR-12652, HST-GO-11136, HST-GO-13855, HST-GO-1331, and STScI Director's Discretionary Research Funds. The W. M. Keck Observatory is operated as a scientific partnership among the California Institute of Technology, the University of California, and NASA. The Observatory was made possible by the generous financial support of the W. M. Keck Foundation.  
\end{acknowledgements}

\bibliographystyle{aa} 
\bibliography{references} 

\begin{appendix}

\section{The odds ratio calculation}
\label{App:Odds Ratio}
To evaluate in a more rigorous way the confidence of the detection we constructed an odds ratio ($\mathrm{OR}$) between the likelihood of the planet ($H_1$) and the false positive ($H_0$) hypothesis. 
Based on the Bayes theorem, it can be calculated as: 
\begin{equation}
\mathrm{OR} = \frac{\mathrm{P}(H_1|x)}{\mathrm{P}(H_0|x)} = \frac{\mathrm{P}(x|H_1,\mu_c)}{\mathrm{P}(x|H_0)} \times \frac{\mathrm{P}(H_1)}{\mathrm{P}(H_0)}
\end{equation}
where $\mathrm{P}(x|H_1,\mu_c)$ and $\mathrm{P}(x|H_0)$ are the likelihood of the data given the signal with flux $\mu_c$ and the likelihood of the data given a false positive, respectively. $\mathrm{P}(H_1)/\mathrm{P}(H_0)$ is the ratio of the priors for $H_1$ and $H_0$.
$\mathrm{P}(x|H_0)$ can be evaluated from the S/N of the bright emission under Gaussian noise assumption. In this case, a 5 sigma detection corresponds to a probability of 0.12\% using the Student t-distribution with 6 degrees of freedom (defined as the number of independent and identically distributed samples at a radial distance $r$, i.e. $(2\pi r/\mathrm{FWHM})-2$). $\mathrm{P}(x|H_1,\mu_c)$ can be estimated after removing the bright emission. There is a 50\% chance for the noise to be higher, and 50\% chance to be lower at that location, meaning $\mathrm{P}(x|H_1,\mu_c)$=0.5. 
To estimate the prior ratio, we considered the expected number of planets in a given mass and separation range around a $\sim$2 $M_{\odot}$ star. Preliminary analysis of the Gemini Planet Imager Exoplanet Survey gives a 6\% probability of 2 $M_{\odot}$ star hosting a planet between 5-13 $M_{\mathrm{Jup}}$ in the 10-100 au separation range \citep{nielsen2017}.
The disk geometry and its L'-band luminosity constrain the mass of the protoplanet around MWC~758 to be $\sim$1--6 $M_{\mathrm{Jup}}$. 
Assuming that planets follow the mass and semi-major axis distributions measured by radial-velocity surveys \citep{cumming2008}, we estimated the likelihood of having a planet between 1-6 $M_{\mathrm{Jup}}$ and between 19-21 au to be $\mathrm{P}(H_0)$= 0.006 (hence, $\mathrm{P}(H_1)=1-\mathrm{P}(H_0)$=0.994). 
Considering that we detected the point-source in two epochs, indicated here by the subscript $a$ and $b$, the odds ratio becomes:

\begin{equation}
\mathrm{OR} = \frac{\mathrm{P}(x_a|H_1,\mu_c)}{\mathrm{P}(x_a|H_0)}\times \frac{\mathrm{P}(x_b|H_1,\mu_c)}{\mathrm{P}(x_b|H_0)} \times \frac{\mathrm{P}(H_1)}{\mathrm{P}(H_0)}=1048.
\end{equation}
The planet hypothesis is thus favored at an odds ratio of 1048:1, providing high confidence that this is, indeed, a true companion.

\section{Reliability of the observed spirals}\label{App:reliability spirals}
\begin{figure*}[h]
\centering
\resizebox{\hsize}{!}{
\includegraphics{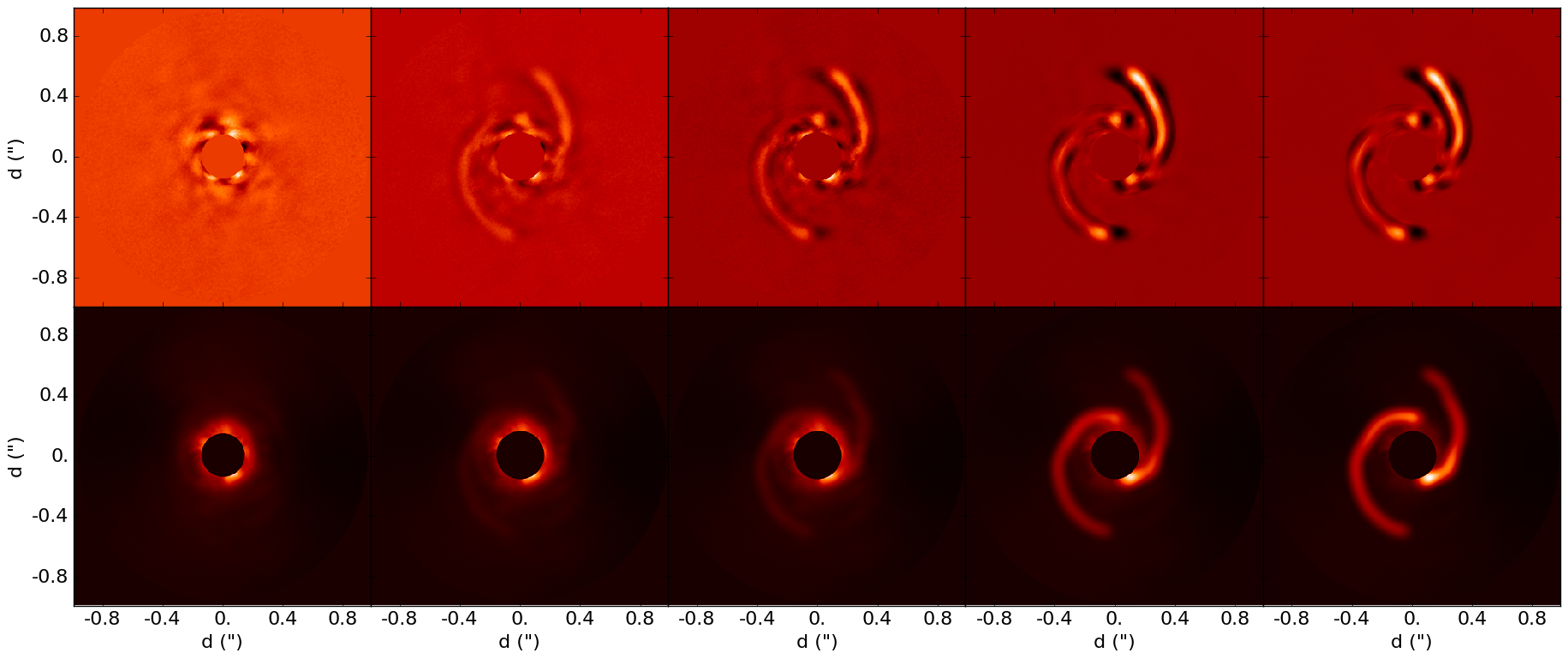}}
  \caption{ADI and RDI comparison on artificial spirals. Top and bottom row show the results of PCA-ADI and PCA-RDI, respectively, on artificially injected spirals. The injected spiral flux increases from left to right.}
     \label{App:test spirals}
\end{figure*}

ADI is known to be an aggressive algorithm that can introduce biases in the results of disk image processing \citep{Milli2012}. We tested the effect of ADI on spiral features, by injecting two artificial spiral arms (similar to S1 and S2) into an ADI cube obtained in similar conditions for a different source showing no disk emission. For this object, a reference star was also observed before and after the target observations to allow for reference star differential imaging \citep[RDI,][]{mawet2013}. In the residual images after PCA analysis, the injected spirals in the ADI data reduction appear to be sharper, but unchanged in shape, compared to RDI (see Figure~\ref{App:test spirals}). No tertiary arm is generated by the injections of the two spiral arms.
These tests give us confidence that the structures detected in the final PCA-ADI images are real.

\section{The source at 2.3"}\label{App:background source}
\begin{figure}
   \centering
   \resizebox{\hsize}{!}
            {
   \includegraphics[height=1.cm]{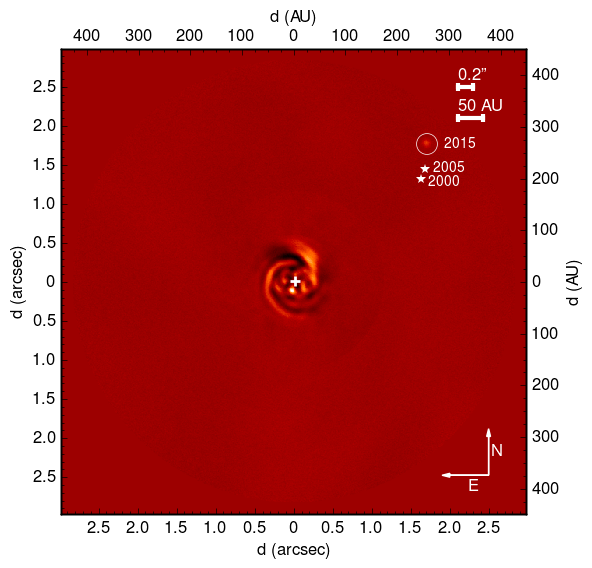}
  }
      \caption{Wide field view of MWC~758 from our L'-band 2015 data. A source is detected at 2\farcs3 from the star. Positions of the same object in the 2000 \citep{grady2005} and 2005 \citep{grady2013} HST observations are plotted on the image, too. North is up and east is toward the left.}
         \label{Figure:background source}
   \end{figure}

After PSF subtraction, a source at a distance $r = 2\farcs315 \pm 0\farcs002$ and $\mathrm{PA}$ = 316\degr$\pm 2\degr$ is recovered in the final 2015 image (see Figure~\ref{Figure:background source}). Its L'-band magnitude is $L'$ = $14.4\pm0.1$ mag.
This source was previously classified as non-comoving \citep{grady2013} and as background object based on its V-magnitude \citep{grady2005}. A re-analysis of the archival STIS and NICMOS data sets \citep{choquet2014}
results in magnitudes $V=17.74 \pm 0.03$ and $J = 15.8 \pm 0.1$ assuming a Kurucz \citep{kurucz1993} A2V stellar model, and $V=18.37 \pm 0.03$ and $J=15.5 \pm 0.1$ for a M2V spectral-type. 
The color information and absolute magnitudes are inconsistent with a young object at the distance of Taurus. They would rather suggests either an M4 main sequence star at the distance of MWC~758, or a red giant much further away. Considering that the probability of an old M dwarf crossing the Taurus star forming region is small, we believe that the source at 2\farcs3 is most likely a background red giant.

\end{appendix}

\end{document}